\documentstyle[12pt,epsf]{article}

\title{Gravitational effects on entangled states 
and interferometer with entangled atoms\thanks{Physics Letters A {\bf 
286}, 102-106 (2001)}}

\author{
Horst von Borzeszkowski$^{{a}}$\thanks{Email:
borzeszk@itp.physik.tu-berlin.de}
\and
Michael B. Mensky$^{{a}{b}}$\thanks{Email: mensky@sci.lebedev.ru}
\\[3pt]$^{{a}}$
Institut f\"ur Theoretische Physik,
Technische Universit\"at Berlin,\\
Hardenbergstr. 36,
D-10623 Berlin, Germany\\
\\[10pt]$^{{b}}$
P.N.Lebedev Physical Institute, 53 Leninskii prosp.,\\
117924 Moscow, Russia\\
}
\date{}

\newcommand{\Fig}[1]{Fig.~\ref{#1}}
\newcommand{\ra}{\rangle}
\newcommand{\st}[1]{{|#1\ra}}

\begin{document}
\maketitle

\begin{abstract}
It is discussed how systems of quantum-correlated (entangled)
particles or atoms behave in external gravitational fields and what
gravitational effects may exist in such systems. An experimental setup
is proposed which improves the sensitivity of the Ramsey interferometer
by the usage of quantum-correlated atoms. Entanglement of $n$ atoms
improves the sensitivity to small phase shifts in $n^2$ times. This
scheme may be used for observing gravity-induced phase shifts in
laboratory.
\end{abstract}

\section{Entangled particles in gravitational fields}

Quantum-correlated, or entangled, states of two or more systems
(for example particles or atoms) proved to be useful for
constructing a new class of quantum devices for
information transfer \cite{EntanglInfo}, quantum cryptography
\cite{EntanglCript} and quantum computing \cite{EntanglComp}. We
shall consider the possibility to use them for observing
gravitational effects.

\subsection{Destruction of correlation by gravity}

It has been shown in the preceding paper \cite{epr-grav} that
correlation of the particles forming a EPR-pair may be violated if the
particles propagate along different trajectories in a gravitational
field. The violation may have the character of distortion (when the
correlation still exists but is altered) or blurring (when the
correlation is partly or completely washed out).

As an example a pair of spin-1/2 particles in correlated spin states
(obtained in the course of the decay of a spin-0 particle) has been
considered in  \cite{epr-grav}. As a result of the correlation, a 
measurement of the spin projection of one of the particles informs the 
experimenter about the spin projection of the other. In case of
spinning particles having been obtained by a decay of a spinless 
particle, their spin projections on the same axis must be oppositely 
directed. 

If external fields are absent, the spin correlation is maintained even 
if the particles move away from each other to any distance. However 
any field which rotate spins, among them gravitational field, 
destructs the correlation or blurs it. The effect is especially 
dramatic for gravitational field since the very concept of space 
direction (along which the spin projections have to be measured) 
cannot be defined globally in this case, but only in each point 
separately \cite{epr-grav}.

\subsection{Violation of spin correlation in gravitational field}

In principle this effect may be used for observing gravitational 
field. If for example all fields except for the gravitational field 
are removed (by screening) but nevertheless the experimenter observes 
that the spin correlation is violated, then the violation is an 
evidence of the presence of a gravitational field.

Correlation may exist and can be experimentally observed (or 
alternatively, its violation may be observed) even in a gravitational 
field where no global definition of space direction (no 
teleparallelism) exists. For this end repeated experiments are needed. 
Arranging the decay of a spin-0 particle many times and each time 
measuring the spin projections of the fragment particles on different 
axes, the experimenter may either find a pair of axes (in two 
different points) such that the corresponding spin projections are 
100\%{} correlated or prove that no such pair of axes exists. In the 
latter case the correlation will be proved to be blurred. In the first 
case it is only distorted. The reason for the distortion is a rotation 
of spins in the gravitational field. One may discover this effect by 
investigating the relation between the two axes (in spatially distant 
points) found in the series of correlation experiments (call them, for 
simplicity, the correlated axes). As a result of this investigation, 
a conclusion about existence and character of the gravitational field 
may be drawn.

Indeed, identifying spatial directions in neighboring points, an
experimenter may, step by step, transfer these directions along any
curve. This is possible, at least, if the gravitational field may be
considered to be static. The result of such transfer may be
mathematically described by the parallel transport of a local basis. 
Thus, two axes in spatially distant point may be put into correspondence 
by the parallel transfer along a certain curve. Let us apply this to 
the axes found in the correlation experiments (``correlated axes''). 
If the parallel transport along some curve (close to a geodesic) 
transfers one of the correlated axes into the other, 
then there is no evidence of a non-vanishing gravitational field. If 
the parallel transport transfers one of the correlated axes into the axis 
which differs from the second correlated axis by some rotation, then 
the conclusion is that a non-vanishing gravitational field is present 
which rotates the particles' spins.

We do not think that the effect of a gravitational field on the 
correlation of spins is directly observable by the present 
experimental technics, so that we do not elaborate the question in 
more detail. Note however that this effect may turn out to be 
important in some specific situations (for example as a part of more 
complicated phenomena in astrophysics) or can become significant due 
to future development of technology.

\subsection{Other gravitational effects based on entanglement}

The correlation of spins is not the only type of entanglement which 
may in principle lead to gravitational effects. Spins may be rotated 
by a gravitational field because and if the rotation group is a 
subgroup of the holonomy group of the given gravitational field. In 
generic case a  holonomy group is a subgroup of the Lorentz group. 
Therefore it may include also boosts (transitions between reference 
frames moving with different velocities). Let two photons be prepared 
in some point in such a way that their frequencies are correlated. 
It may happen that each of the frequencies has an undetermined value 
but the frequencies of the two photons are correlated (in the sense of 
quantum correlation, i.e., by entanglement). For example, the energies 
of the separate photons may be indefinite but the total energy is fixed.  
If then the photons are traveling into another point along different 
paths in a gravitational field, the correlation between their 
frequencies may be violated (distorted or blurred) when the photons 
arrive at the final point. This is nothing else than the effect of the 
gravitational red shift but now considered for a pair of entangled 
photons.

Moreover, in principle gravitational effects connected with the affine 
holonomy group may also be observed. This brings also the translation 
group into play, besides the Lorentz group. Therefore, gravitationally 
induced time and/or space shifts may be observed with the help of a 
pair of quantum-correlated particles (or even with a greater number of 
correlated particles).

\section{Detection of gravity by an interferometer with 
entangled atoms}

Now let us consider ordinary schemes of observing gravitational 
effects which include no quantum correlation and show that even in 
such schemes the effect may be increased by quantum correlation. We 
shall present a scheme which makes use of quantum correlation to 
increase the gravity-induced phase shift. More precisely, it will be 
shown that the quantum correlation of two or more atoms may be used to 
add up the phase shifts of singular atoms in Ramsey interferometer.

\subsection{Idea of an interferometer with entanglement}

\begin{figure}[ht]
\let\picnaturalsize=N
\def\picsize{7cm}
\ifx\nopictures Y\else{\ifx\epsfloaded Y\else\input epsf \fi
\let\epsfloaded=Y
{\hspace*{\fill}
 \parbox{10cm}{\ifx\picnaturalsize N\epsfxsize \picsize\fi
                           \epsfbox{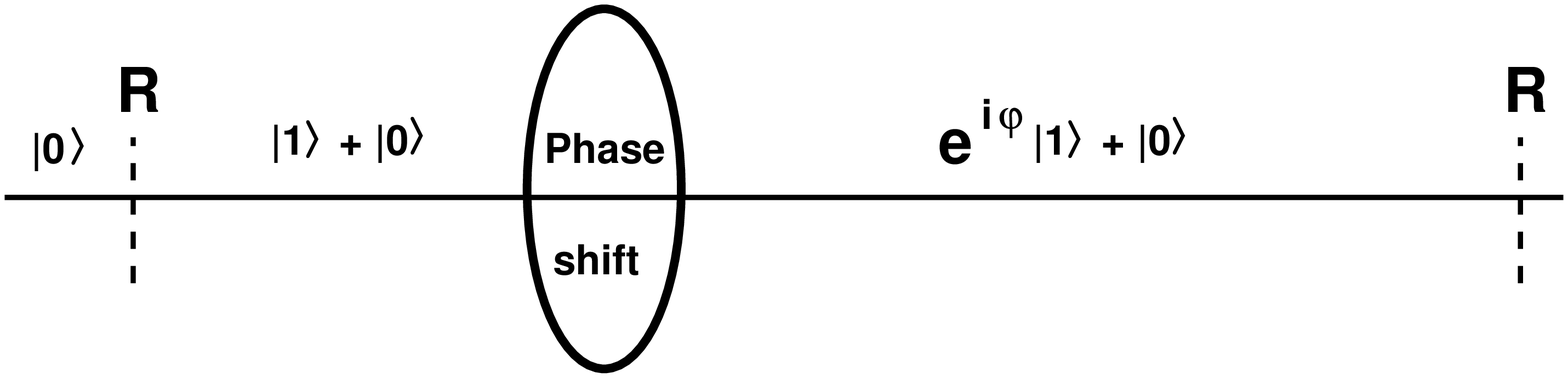}}\hspace*{\fill}}\fi\\[2ex]
{\hspace*{\fill}
 \parbox{10cm}{\ifx\picnaturalsize N\epsfxsize \picsize\fi
                               \epsfbox{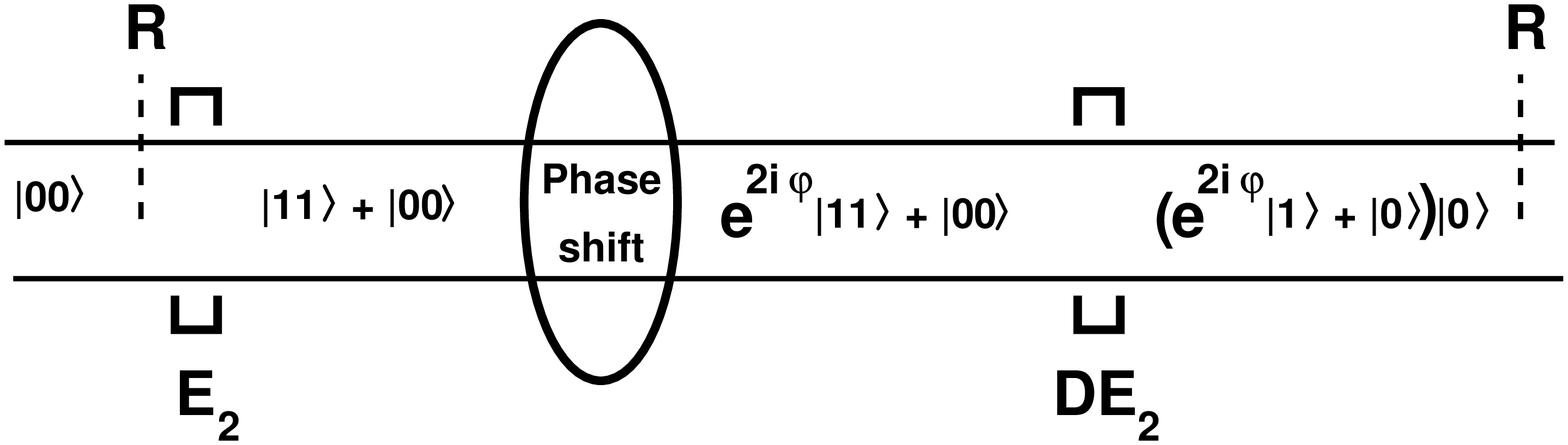}}\hspace*{\fill}}\\
{\hspace*{\fill}
 \parbox{10cm}{\ifx\picnaturalsize N\epsfxsize \picsize\fi
                                 \epsfbox{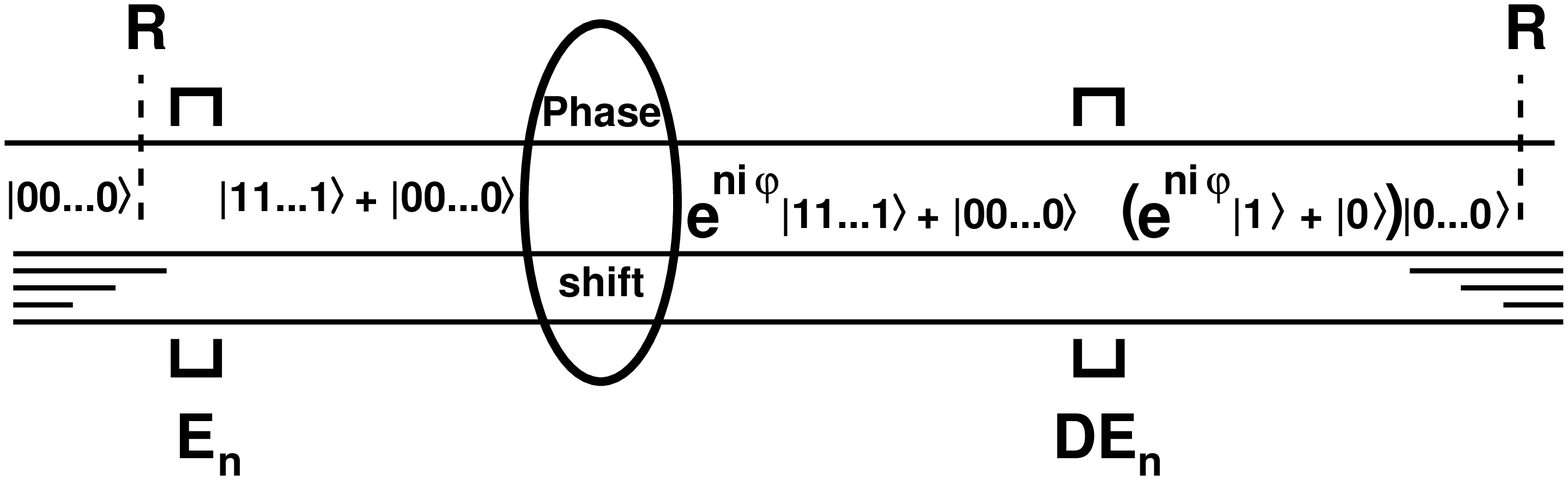}}\hspace*{\fill}}}
\caption{Ramsey interferometer with one atom (upper diagram), two 
entangled atoms (middle) and $n$ entangled atoms (lower). Rabi pulses 
${\rm R}$ are applied only to the first atom, but the phase shift is 
determined by state differences of all entangled atoms. Device ${\rm 
E}_n$ realizes an entanglement of the atoms while ${\rm DE}_n$ does 
disentangle them. The probability for the first particle to transit 
from the initial state $\st{0}$ to the final state $\st{1}$ depends on 
the total phase shift which is proportional to the number of entangled 
atoms. The sensitivity of the interferometer to small phase shifts is 
thus increased by the factor $n^2$.}
\label{FigRamsey}
\end{figure}

The idea is as follows (see Fig.~\ref{FigRamsey}).
Owing to the action of an external, for example gravitational field 
one energy level of the atom acquires a phase shift with respect to 
the other level \cite{RamsGrav}.\footnote{The reason for the phase 
shift is that the path the atom is traveling along depends on its 
state.} This phase shift may be detected with the help of the Ramsey 
interferometer (Fig.~\ref{FigRamsey}, upper diagram). If we have 
instead of this a quantum-correlated (entangled) state of two 
(Fig.~\ref{FigRamsey}, middle diagram) or many (Fig.~\ref{FigRamsey}, 
lower diagram) atoms, then the phase shift is twice or many times 
greater. Therefore, the sensitivity of the interferometer is 
increased.

Notice that an analogous proposal for the Ramsey spectrometer
instead of Ramsey interferometer, which has been put forward in
\cite{RamsSpectr}, fails. In the case of a spectrometer the deviation of the 
frequency of the Rabi pulse from the resonance frequency of the atom 
is measured. In \cite{RamsSpectr} it was suggested to improve the 
sensitivity of the spectrometer making use of entangled atoms. However 
just in this case the idea does not work. The reason is that the 
Ramsey interference in the spectrometer is owing to the phase 
difference of the two Rabi pulses which is determined by the flying 
time of the atom between the two Rabi pulses. In this case the phase 
shift does not depend on the number of the entangled particles 
(Eqs.~(5,6,8) and therefore most of the rest formulas in 
\cite{RamsSpectr} contain mistakes).

\subsection{Scheme of an interferometer with entanglement}

Contrary to the case of the spectrometer, the sensitivity of the 
interferometer will be actually improved if the interfering atoms are 
entangled. A concrete scheme of the interferometer with two entangled 
atoms may be worked out in the following way.

\begin{figure}[ht]
\let\picnaturalsize=N
\def\picsize{6cm}
\ifx\nopictures Y\else{\ifx\epsfloaded Y\else\input epsf \fi
\let\epsfloaded=Y
{\hspace*{\fill}
 \parbox{14cm}{\ifx\picnaturalsize N\epsfxsize \picsize\fi
                           \epsfbox{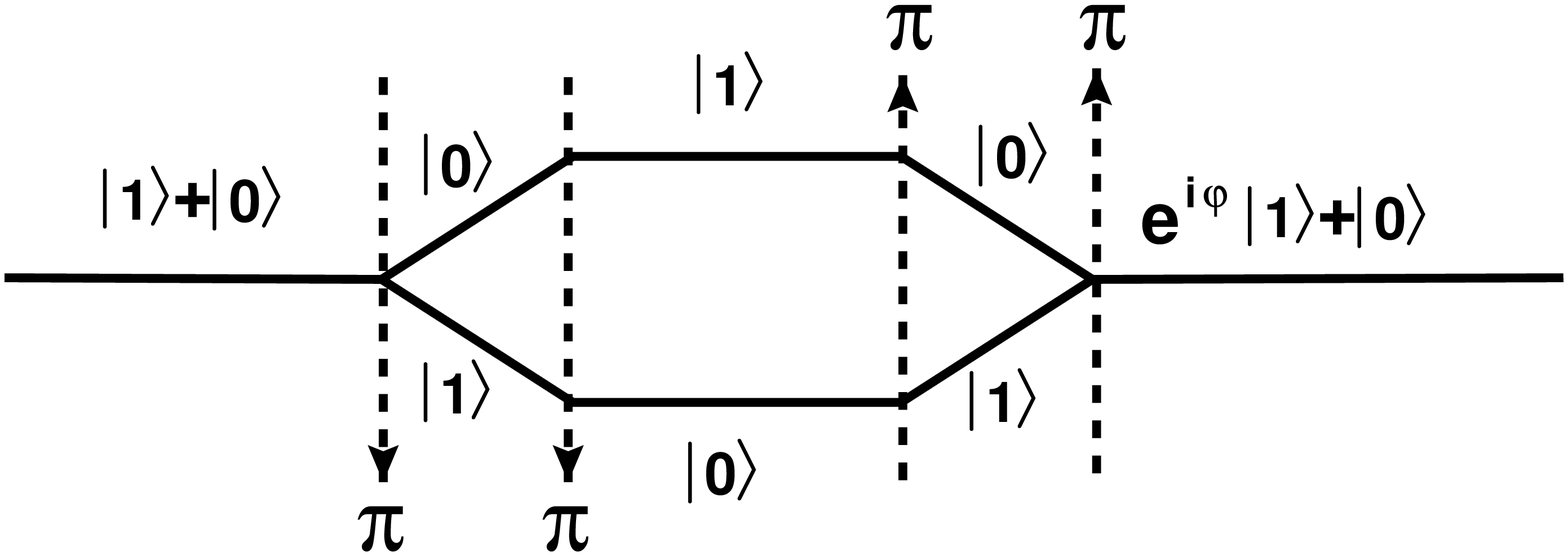}}\hspace*{\fill}}\fi\\[2ex]
{\hspace*{\fill}
 \parbox{14cm}{\ifx\picnaturalsize N\epsfxsize \picsize\fi
                                 \epsfbox{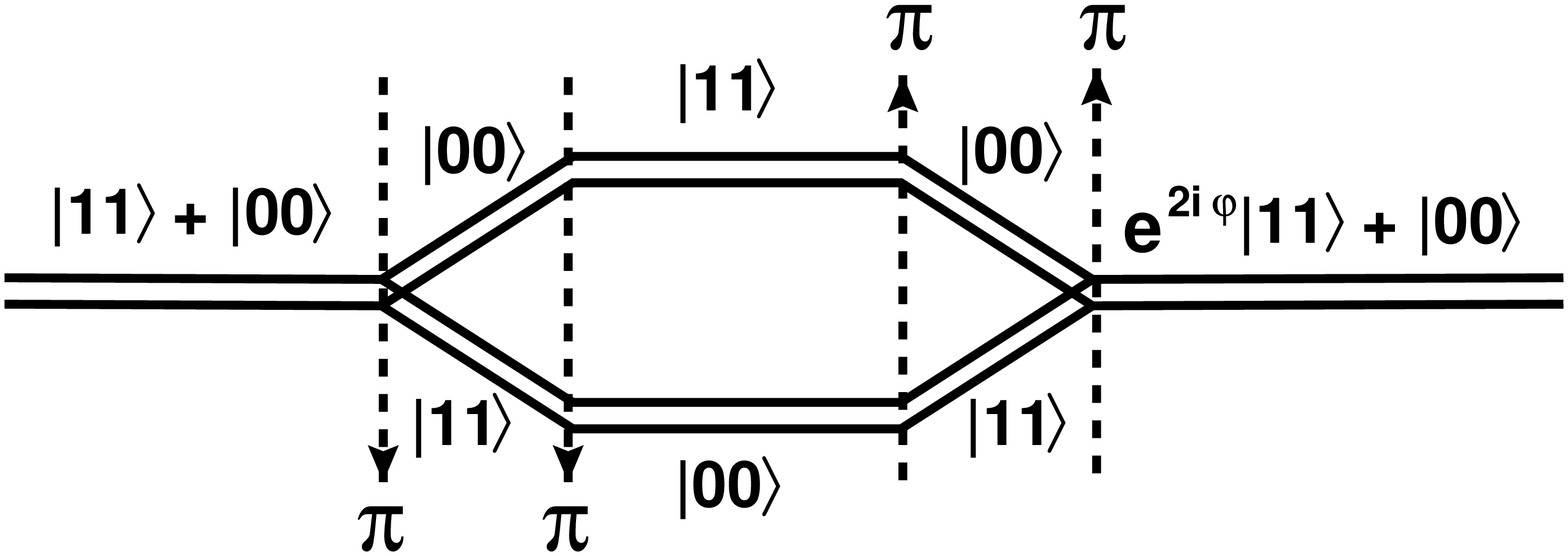}}\hspace*{\fill}}}
\caption{Atom interferometer without (upper diagram) and with
(down) entanglement.}
\label{FigPh-shift}
\end{figure}

The role of entanglement is easily seen from the comparison of the 
two diagrams of \Fig{FigPh-shift}.
The experimental setup is such that atoms in the states $\st{0}$, 
$\st{1}$ move in a gravitational field along one of two paths 
depending on the state of the given atom. This is achieved by the 
usual method \cite{RamsGrav}, i.e. with the help of $\pi$-pulses of 
resonance radiation in the form of running waves (the direction of the 
running wave is pointed out by the arrow). Acting on the ground state 
$\st{0}$ this radiation causes the atom to absorb one photon. As a result, 
the ground state converts into the excited state $\st{1}$. The 
transversal linear momentum of the atom, initially null, becomes equal 
to the linear momentum $k$ of the absorbed photon (directed along the 
direction of the running wave). On the contrary, the atom which is 
initially in the state $\st{1}$ issues the same photon and therefore 
transits into the state $\st{0}$ with the linear momentum $-k$. Thus, 
during the next period of time, the direction of motion of the atom 
depends on its state as it is drawn in \Fig{FigPh-shift}. Then, after 
the action of the second $\pi$-pulse (with the same direction of 
running wave), the atoms return to the initial states with the initial 
null transversal linear momentum. Now each atom is moving along one of 
two parallel lines depending on its state. Then the pair of 
$\pi$-pulses with the opposite direction of running waves brings the 
atom together again as is shown in \Fig{FigPh-shift}.

Let the passage of the states $\st{0}$, $\st{1}$ along different paths 
in a gravitational field (as in \Fig{FigPh-shift}) give the phase 
factor $\exp(i\varphi)$ only for the state $\st{1}$. Then the 2-atom 
states $\st{11}=\st{1}\st{1}$, $\st{00}=\st{0}\st{0}$, when passing 
through the same device, will differ by the factor $\exp(2i\varphi)$ 
(\Fig{FigPh-shift}, lower diagram). The $n$-atom state $\st{11\dots 
1}$ will acquire the factor $\exp(ni\varphi)$ as 
compared to the state $\st{00\dots 0}$ . Therefore, the phase shift 
will be $n$ times greater for $n$ entangled atoms than in the case of 
not-entangled atoms. The sensitivity to the phase shift will be 
improved correspondingly. 

\begin{figure}[ht]
\let\picnaturalsize=N
\def\picsize{7cm}
\def\picfilename{detect.eps}
\ifx\nopictures Y\else{\ifx\epsfloaded Y\else\input epsf \fi
\let\epsfloaded=Y
\centerline{\ifx\picnaturalsize N\epsfxsize \picsize\fi \epsfbox{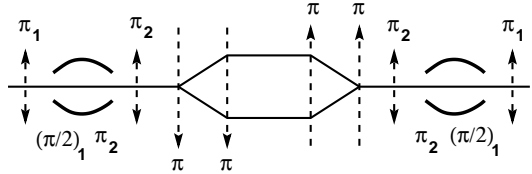}}}\fi
\caption{Scheme of an interferometer with two entangled atoms}
\label{FigDetect}
\end{figure}

We have still to point out a possible construction of devices 
${\rm E}_n$ and ${\rm DE}_n$ of \Fig{FigRamsey}. In other words, 
we have to explain how the initial entangled state may be formed 
and how the final entangled state is converted to the disentangled one 
so that the phase shift may be observed experimentally. To this end, 
consider a scheme which realizes this task in the case $n=2$. The 
entanglement and disentanglement of atoms may be achieved by passing 
them through the resonance microcavities as it is shown in 
\Fig{FigDetect}.
The central part of the figure is identical to the lower diagram of 
\Fig{FigPh-shift}. The first (left) part of \Fig{FigDetect} (device 
${\rm E}_2$) realizes the entanglement of the atoms and the last (right) 
part (device ${\rm DE}_2$) again disentangles them.

Two atoms entering the device ${\rm E}_2$ are in the ground state
$\st{0}$ so that the 2-atom state is $\st{00}$. Then the first
atom is subject to the $\pi$-pulse of the resonance standing-wave
radiation (in \Fig{FigDetect} this is denoted by $\pi_1$). This gives
state $\st{10}$. Then the first atom is flying through a microcavity 
during the time corresponding to the $\pi/2$-pulse (in the figure this is
denoted by $(\pi/2)_1$). The initial state of the cavity is vacuum
$\st{0}_\gamma$ so that the initial state of both atoms and the
microcavity is $\st{1}_1\st{0}_\gamma\st{0}_2$. After the cavity is
passed by the first atom, this state converts into the superposition
$(\st{1}_1\st{0}_\gamma-\st{0}_1\st{1}_\gamma)\st{0}_2/\sqrt{2}$ (a
superposition of the initial state and the state resulting from 
radiation of a photon by the first atom into the cavity). After this 
the second atom is flying through the same cavity during the time period 
corresponding to the $\pi$-pulse (in the figure this is denoted by 
$\pi_2$ under the cavity). This gives 
$(\st{1}_1\st{0}_2-\st{0}_1\st{1}_2)\st{0}_\gamma/\sqrt{2}$, thus the 
cavity is finally in the vacuum state as it has been initially. At last, 
the second atom is subject to the $\pi$-pulse of resonance radiation. 
This results in the state 
$(\st{1}_1\st{1}_2+\st{0}_1\st{0}_2)/\sqrt{2}$ of the atoms. Thus the 
entangled state $(\st{00}+\st{11})/\sqrt{2}$ is formed as the output 
of ${\rm E}_2$.

The action of the disentangling device ${\rm DE}_2$ is analogous, but with
the opposite order of all operations. Notice that the second atom has
to pass the second microcavity (included in ${\rm DE}_2$) before the first
one does so. This may be provided by a greater velocity of the second atom
on its way from the first cavity to the second one.

If there is no phase shift between the states $\st{11}$ and $\st{00}$ 
traveling along different paths, then the state $\st{11}+\st{00}$ 
entering the device ${\rm DE}_2$ is converted into $\st{00}$. If however 
the state $\st{11}$ obtains some phase shift $2\varphi$ during its travel 
in the gravitational field so that the state entering ${\rm DE}_2$ is 
$e^{2i\varphi}\st{11}+\st{00}$ then the output state is a mixture of 
$\st{00}$ and $\st{10}$. The probability that the state $\st{10}$ will 
be detected at the output is equal to 
$$ 
{\rm Prob}(\st{10})= \frac{1}{2}(1-\cos2\varphi) 
$$ 
which is equal to $\varphi^2$ for small $\varphi$. For comparison, in 
the interferometer using no entanglement (as in the upper diagram of 
\Fig{FigPh-shift}) the probability to discover the phase shift 
$\varphi$ would be equal to $(1-\cos\varphi)/2$ which is equal to 
$\varphi^2/4$ for small $\varphi$.

Thus, entanglement of two atoms increases the probability to observe
the phase shift by the factor 4. If $n$ atoms are entangled, the
probability will be $n^2$ times greater than it is in the scheme 
without entanglement. An experimental setup of the type presented in 
\Fig{FigDetect} is feasible with the present technology.

\vspace{5mm}

\centerline{\bf ACKNOWLEDGEMENT}

The work was supported in part by 
the Deutsche Forschungsgemeinschaft, grant 436 RUS 17/12/00.

\end{document}